\documentclass[journal,twoside,web]{ieeecolor}
\usepackage{generic}
\usepackage{cite}
\usepackage{amsmath,amssymb,amsfonts}
\usepackage{algorithmic}
\usepackage{graphicx}
\usepackage{algorithm,algorithmic}
\usepackage{multirow}
\usepackage{textcomp}
\usepackage{generic}
\usepackage{cite}
\usepackage{amsmath,amssymb,amsfonts}
\usepackage{algorithmic}
\usepackage{graphicx}
\usepackage{textcomp}
\usepackage{multirow}
\usepackage{array}
\usepackage{makecell}
\usepackage{color}
\usepackage{soul}
\usepackage[export]{adjustbox} 
\usepackage{bm}
\usepackage{tabularx}
\usepackage{caption}
\usepackage{booktabs} 
\soulregister\cite7 
\soulregister\citep7 
\soulregister\citet7 
\soulregister\ref7 
\soulregister\pageref7 
\makeatletter
\let\NAT@parse\undefined
\makeatother
\usepackage[colorlinks=true,
           linkcolor=blue,
           anchorcolor=blue,
           citecolor=green,
           urlcolor=blue,
           ]{hyperref}
\def\BibTeX{{\rm B\kern-.05em{\sc i\kern-.025em b}\kern-.08em
    T\kern-.1667em\lower.7ex\hbox{E}\kern-.125emX}}
\begin{document}
\title{SCKansformer: Fine-Grained Classification of Bone Marrow Cells via Kansformer Backbone and Hierarchical Attention Mechanisms}
\author{Yifei Chen, Zhu Zhu, Shenghao Zhu, Linwei Qiu, Binfeng Zou, Fan Jia, Yunpeng Zhu, Chenyan Zhang, Zhaojie Fang, Feiwei Qin, Jin Fan, Changmiao Wang, Gang Yu, Yu Gao
\thanks{This work was supported by National Key Research and Development Program of China (No. 2023YFC2706400), GuangDong Basic and Applied Basic Research Foundation (No. 2022A1515110570), Innovation Teams of Youth Innovation in Science and Technology of High Education Institutions of Shandong Province (No. 2021KJ088), Natural Science Foundation of Zhejiang Province (No. LY21F020015), the Open Project Program of the State Key Laboratory of CAD\&CG (No. A2304), Zhejiang University and National Undergraduate Training Program for Innovation and Entrepreneurship (No. 202310336074). (Yifei Chen and Zhu Zhu contributed equally to this work.) (Corresponding author: Zhu Zhu, Feiwei Qin.) }
\thanks{Y. Chen, S. Zhu, B. Zou, F. Jia, Y. Zhu, C. Zhang, Z. Fang, F. Qin and J. Fan are with Hangzhou Dianzi University, Hangzhou, 310018, China (e-mail: (chenyifei, 22320220, 21321333, 18120401, 22330634, 21321231, 21321206, qinfeiwei, fanjin)@hdu.edu.cn).}
\thanks{Z. Zhu and G. Yu are with Children's Hospital, Zhejiang University School of Medicine, Hangzhou, 310000, China (e-mail: zhuzhu$\_$cs@zju.edu.cn; yugbme@zju.edu.cn).}
\thanks{L. Qiu is with Beihang University, Beijing, 100191, China (e-mail: qiulinwei@buaa.edu.cn).}
\thanks{C. Wang is with Shenzhen Research Institute of Big Data, Shenzhen, 518172, China (e-mail: cmwangalbert@gmail.com).}
\thanks{Y. Gao is with Zhejiang Hospital, Hangzhou, 310030, China (e-mail: 84853106@qq.com).}}

\maketitle

\begin{abstract}
The incidence and mortality rates of malignant tumors, such as acute leukemia, have risen significantly. Clinically, hospitals rely on cytological examination of peripheral blood and bone marrow smears to diagnose malignant tumors, with accurate blood cell counting being crucial. Existing automated methods face challenges such as low feature expression capability, poor interpretability, and redundant feature extraction when processing high-dimensional microimage data. We propose a novel fine-grained classification model, SCKansformer, for bone marrow blood cells, which addresses these challenges and enhances classification accuracy and efficiency. The model integrates the Kansformer Encoder, SCConv Encoder, and Global-Local Attention Encoder. The Kansformer Encoder replaces the traditional MLP layer with the KAN, improving nonlinear feature representation and interpretability. The SCConv Encoder, with its Spatial and Channel Reconstruction Units, enhances feature representation and reduces redundancy. The Global-Local Attention Encoder combines Multi-head Self-Attention with a Local Part module to capture both global and local features. We validated our model using the Bone Marrow Blood Cell Fine-Grained Classification Dataset (BMCD-FGCD), comprising over 10,000 samples and nearly 40 classifications, developed with a partner hospital. Comparative experiments on our private dataset, as well as the publicly available PBC and ALL-IDB datasets, demonstrate that SCKansformer outperforms both typical and advanced microcell classification methods across all datasets. Our source code and private BMCD-FGCD dataset are available at \href{https://github.com/JustlfC03/SCKansformer}{https://github.com/JustlfC03/SCKansformer}.
\end{abstract}


\section{Introduction}
\label{sec:introduction}
\IEEEPARstart{I}{n} recent years, the morbidity and mortality rates of malignant tumors, such as acute leukemia, have shown a clear upward trend worldwide \cite{c19chhikara2023global}. In the diagnostic process, hospitals initially employ cytological examinations of peripheral blood smears and bone marrow smears to determine the specific type of malignant tumor. For instance, based on cytology examination results, doctors can identify the type of leukemia and formulate targeted treatment plans. In cytological examinations, medical personnel focus on the number and proportion of various types of white blood cells in a sample from a leukemia patient compared to a healthy individual, relying on the percentage of primitive cells to determine the type and progression of the disease. Thus, the accuracy of bone marrow or peripheral blood smear tests hinges on the ability to accurately count the different types of blood cells in the sample. The results of bone marrow smear cytology are especially critical for clinicians as they serve as essential references for diagnosing and treating the disease.

Within a healthy body, the number of white blood cells remains within a specific range. However, the number and proportion of white blood cells can fluctuate in response to inflammation or blood disorders. Therefore, routine blood tests are a vital tool for assessing overall health and assisting doctors in diagnosis \cite{c20brinati2020detection}. For example, a sudden increase in lymphocytes may indicate conditions such as whooping cough, a bacterial infection, or cancer of the lymphatic system. Although initial blood tests can provide direction for diagnosing a disease by evaluating the number and percentage of different white blood cells, they cannot definitively diagnose a disease. Consequently, peripheral blood smears are often necessary for accurate diagnosis and treatment. This cost-effective diagnostic method is generally employed when routine blood tests yield abnormal results and is designed to detect morphological changes in all types of blood cells in peripheral blood. By examining the morphology and number of white blood cells, red blood cells, and platelets, one can gain a better understanding of the underlying cause of a disease.

Clinicians rely on abnormalities in blood cell morphology for comprehensive diagnosis, such as identifying primitive or naive blood cells, toxic granules, vacuolar degeneration in neutrophils, and abnormal erythrocyte morphology (e.g., spherical, oval, or echinocytic erythrocytes). These observations are crucial for diagnosing and monitoring diseases in clinical patients. For precise diagnosis and treatment, especially for high-risk diseases, peripheral blood smears alone are insufficient. Instead, bone marrow smear cytomorphology is necessary. This method is primarily used to diagnose hematopoietic diseases such as leukemia, aplastic anemia, multiple myeloma, and other major malignant conditions \cite{c34gars2018bone}. Bone marrow smear cytology involves using a microscope or automated instrument to examine the morphology and number of blood cells in the bone marrow. The density of nucleated cells or the ratio of nucleated cells to mature red blood cells in the bone marrow smear is used to estimate the degree of proliferation of nucleated cells. Since the bone marrow is the body's primary blood-forming organ, this method needs to examine up to nearly 40 different types of white blood cells, unlike peripheral blood smears, which typically distinguish only three to five types of white blood cells.

Currently, hospitals typically use hematology analyzers to count blood cells. These devices primarily rely on physical or chemical solutions for counting, but they have limitations and can only identify a few types of blood cells. A comprehensive bone marrow smear cytomorphology, covering forty kinds of blood cells, is usually required to diagnose major risk diseases \cite{c33bain2001bone}. Therefore, relying solely on a hematology analyzer is insufficient for clinicians to diagnose primary risk diseases accurately. Additionally, blood analyzers are limited to counting cells and do not provide microscopic images for further clinical evaluation. Manual counting, on the other hand, is time-consuming and labor-intensive, and results may vary between doctors with different levels of clinical experience and cognition, leading to inconsistent diagnostic standards. Consequently, improving the automation of Hematology Microscopic Examination is crucial for better assisting physicians in diagnosing significant blood diseases.

With the rapid advancement of computer hardware, deep learning has made significant strides in computer vision. Increasingly, researchers are adopting deep learning models to study blood cell classification. Compared to traditional machine learning methods, deep learning models simplify the processes of image processing and feature extraction, making them more automated. However, cell examination in bone marrow blood microscopy images remains a complex challenge \cite{c22matek2021highly}. The accurate classification of multiple similar classes of blood cells is particularly difficult due to the large number of cell types and the presence of similar features among different cell classes.

Moreover, compared to image recognition in natural settings, blood cell classification for major blood diseases, such as leukemia diagnosis, faces several challenges due to the need for bone marrow picture cytology:
\begin{enumerate}
    \item \textbf{Variability in Imaging Conditions:} Different hospitals, mechanical equipment, and imaging environments produce bone marrow blood cell images with varying colors, making accurate identification challenging.
    \item \textbf{Long-Tail Distribution:} Due to the low incidence of major risk diseases, bone marrow blood cell datasets often exhibit a long-tail problem, with minimal sample distribution for rare disease categories and substantial sample distribution for common disease categories, leading to highly heterogeneous category distributions.
    \item \textbf{Resolution and Contrast Issues:} Unlike natural images obtained in sunlight, microscopic images of bone marrow blood samples are captured using specialized imaging equipment and processing techniques. These images typically exhibit lower resolution and contrast, resulting in fewer discernible features compared to natural images. Consequently, microscopic images are challenging for models to recognize.

    \item \textbf{Minimal Differences Between Cell Types:} The differences between various types of bone marrow blood cells in smear cytology are often minimal. For example, distinguishing between early erythroblasts, intermediate erythroblasts, and late erythroblasts is particularly difficult due to their subtle differences.
  \end{enumerate}

To address these challenges of automatic classification and recognition of blood cells in bone marrow microscopic images, this paper proposes a fine-grained classification model for bone marrow blood cells, SCKansformer. 
This method not only provides essential reference standards for clinicians to diagnose and treat major dangerous diseases but also further enhances clinicians' ability to diagnose acute leukemia and other significant hematological diseases, thus holding substantial significance in clinical auxiliary diagnosis. The primary contributions of this paper are as follows:
\begin{itemize}
    \item \textbf{Kansformer Encoder:} We propose the Kansformer Encoder, which utilizes the Kolmogorov-Arnold Network instead of the traditional Multilayer Perceptron layer within the Transformer Encoder. This substitution addresses the shortcomings in handling high-dimensional image data, such as micrographs, enhancing the nonlinear feature representation and interpretability of the model, thereby further improving the classification accuracy of bone marrow blood cells.

    \item \textbf{Enhanced Feature Representation:} Our SCKansformer model incorporates the SCConv Encoder and the Global-Local Attention Encoder. The SCConv Encoder reduces feature redundancy through the Spatial Reconstruction Unit and Channel Reconstruction Unit, optimizing the feature representation of target cells. The Global-Local Attention Encoder combines a Multi-head Self-Attention module with a Local Part module to effectively capture both global and local features of microscopic images, thus improving classification reliability.
    
    \item \textbf{BMCD-FGCD Dataset:} Given the challenges in collecting and labeling bone marrow blood sample images, there is a lack of accurately labeled, high-resolution, large-scale bone marrow blood cell datasets. In collaboration with the Department of Hematology at Zhejiang Hospital in Hangzhou, Zhejiang Province, our team has established the Bone Marrow Cell Dataset for Fine-Grained Classification (BMCD-FGCD), containing over 10,000 data points across nearly forty classifications. We have made our private BMCD-FGCD dataset available to other researchers, contributing to the field's advancement.
    
    \item \textbf{Experimental Validation:} We conducted comparative experiments using the private BMCD-FGCD dataset and the public PBC and ALL-IDB datasets to benchmark the performance of advanced classification methods. The results unequivocally demonstrate that the SCKansformer model outperforms other methods across all datasets. Notably, the SCKansformer model effectively manages imbalanced cell class datasets while maintaining superior performance on balanced cell class datasets, highlighting its broad applicability. Furthermore, ablation experiments were carried out to verify the effectiveness and necessity of each module within the SCKansformer model, thereby reinforcing its comprehensive utility.
\end{itemize}

\section{Related Work}
\subsection{Traditional Clinical Blood Cell Classification Methods}

Early classification of white blood cells was predominantly conducted by trained healthcare professionals using microscopes to observe the color, shape, size, and texture of the cells for identification, classification, and counting. This manual method was plagued by low accuracy, high workloads, subjective bias, and inefficiency. In the Mid-20th century, Wallace H. Coulter et al. introduced the Coulter Principle \cite{c17berkson1935laboratory}. This principle posits that when an electrolyte flows through a tiny aperture, the suspended particles within the electrolyte will also flow through, displacing the electrolyte of the same volume. Due to the constant voltage in the circuit, this displacement causes a change in the resistance across the aperture, resulting in potential pulses. The strength and frequency of these pulses are directly proportional to the volume and number of suspended particles. This principle provided a novel method for measuring the volume of microparticles, thus enabling the counting of cell classes and becoming the gold standard for cell counting.

Subsequently, three- and five-category hematology analyzers based on the Coulter principle were developed and quickly gained widespread adoption, gradually replacing traditional manual examination. Following this, automated electrical or optical cell sorting and counting methods, utilizing flow cytometry, further enhanced processing speed and accuracy. 

While these physicochemical-based leukocyte analyzers provide accurate classifications for clinical examinations, they are less effective in assessing abnormal leukocyte morphology. Additionally, they require expensive medical devices and highly specialized personnel, which limits their utility in routine leukocyte research.

\subsection{Machine Learning Blood Cell Classification Methods}

In the 1980s, numerous researchers began exploring the use of computer image processing techniques and traditional machine learning algorithms to simulate the analysis of leukemia cell morphology under a microscope. Researchers employed various methods to process images of white blood cells. For instance, Ramoser et al. \cite{c1ramoser2006leukocyte} initially applied K-means clustering to identify the nucleus location. They then used the MSER algorithm to threshold the image at different levels, followed by further differentiation of the nucleus and cytoplasm using K-means clustering. Based on the segmented leukocytes, they extracted color and shape features and subsequently classified the leukocytes using a Support Vector Machine (SVM). Similarly, Duan et al. \cite{c2duan2019leukocyte} utilized the SMACC algorithm to extract leukocyte regions and morphological operations to determine leukocyte locations. They employed the ISODATA clustering algorithm for further segmentation to identify the nucleus region. Based on this segmentation, they also tried to extract spectral, texture, and shape features and classified the leukocytes using an SVM.

Addressing the challenges posed by noise and staining differences in leukocyte images, some researchers focused on image preprocessing to remove noise or enhance image quality. Bikhet et al. \cite{c3bikhet2000segmentation} employed median filtering to eliminate noise, followed by threshold segmentation to separate leukocytes from the background, and edge detection to extract ten distinguishing features. Prinyakupt et al. \cite{c4prinyakupt2015segmentation} proposed a linear and Bayesian classifier based on leukocyte segmentation. They enhanced the leukocyte image using the red and blue channels, applied threshold segmentation to detect the nucleus, and extracted 15 features for classification using linear and Bayesian classifiers. Pavithra et al. \cite{c5pavithra2015white} converted images to grayscale, applied the Sobel operator for edge detection, performed image smoothing using median filtering, and used a marker-controlled watershed algorithm to avoid over-segmentation. Morphological operations were then employed to obtain accurate leukocyte locations for counting. Mohapatra et al. \cite{c6mohapatra2014ensemble} enhanced image contrast, utilized the shaded C-mean clustering method for segmentation, and extracted features based on the segmented leukocytes. They performed feature learning using an integrated classifier.

Given the high dimensionality and inter-correlation of extracted features, which led to spatial instability and weak generalization, researchers often reduced feature dimensionality to improve model performance. Huang et al. \cite{c7huang2012computer} used threshold segmentation to locate cell nuclei, extracted 80 texture features and five shape features, and employed Principal Component Analysis (PCA) to identify relevant features for distinguishing five types of leukocytes. They then applied a genetic algorithm-based K-means classifier for automatic leukocyte classification. Nazlibilek et al. \cite{c8nazlibilek2014automatic} also used threshold segmentation for leukocyte identification, followed by PCA to extract optimal features, and classified leukocytes with 95\% accuracy using a Multilayer Perceptron (MLP).



\subsection{Neural Network Blood Cell Classification Methods}

The advancement of computer processing power and the emergence of deep neural networks have significantly promoted the research of leukocyte classification methods. Habibzadeh et al. \cite{c9habibzadeh2013white} compared Convolutional Neural Networks (CNNs) with commonly used SVM classifiers and demonstrated that CNNs could accurately distinguish leukocyte classes even in low-quality blood images. To improve classification accuracy, Ma Li et al. \cite{c13jha2019mutual} replaced the Batch Normalization (BN) residual block in ResNet with the Instance BN residual block and utilized an improved loss function to filter out easily classifiable samples. This adjustment allowed the model to focus on learning from difficult-to-classify samples, achieving an accuracy of 92\% on the BCCD dataset, marking a significant improvement in leukocyte classification.

Despite these advancements, the challenge of accurately classifying leukocytes in blood images persisted, especially given that leukocytes constitute a relatively small portion of these images. To address this, Choi et al. \cite{c10choi2017white} proposed a two-stage CNN model incorporating both local and global models. The global model performed leukocyte dichotomization, while the local model further classified the leukocytes, achieving 97.06\% accuracy on the decile dataset. Although CNNs can directly process blood image data, leukocytes constitute a relatively small portion of these images. Consequently, some researchers have employed image segmentation to crop leukocyte images before using CNNs for feature learning.

However, recognizing the need for precise localization of leukocytes, Banik et al. \cite{c11banik2020automatic}. They applied color space transformation and the K-means algorithm to segment leukocyte nuclei, localized the leukocytes, cropped them from the blood images, and proposed a CNN model that fused features from the first and last layers for classification. The quest for more effective leukocyte classification methods continued with Novoselnik et al. \cite{c12novoselnik2018automatic}, who extracted blue channel data from blood samples, used threshold segmentation to locate leukocytes, cropped the leukocytes, and classified them using LeNet-5, achieving 81.11\% accuracy. This highlighted the potential of integrating traditional image processing techniques with CNNs.

Further pushing the boundaries, Jha et al. \cite{c13jha2019mutual} used a hybrid model based on mutual information to segment leukocytes, combining the results of the active contour model and the fuzzy C-means algorithm. They then extracted features from the segmented images and trained a CNN classifier using the temporal Sine Cosine Algorithm. At the same time, Unlike previous traditional image segmentation methods, some researchers have employed deep learning-based semantic segmentation techniques to locate leukocyte images. Reena et al. \cite{c14reena2020localization} used DeepLabv3+ for segmentation and subsequently used the AlexNet model for classification, achieving an accuracy rate of 98.87\%. Tsykunov et al. \cite{c15tsykunov2020application} designed a two-stage leukocyte classification model, first using U-Net for leukocyte location detection and then employing ResNet-50 for classification.

However, when diagnosing major blood disorders such as leukemia, clinicians also rely on bone marrow smear cytomorphology. Unlike peripheral blood smear examinations, which distinguish between three to five types of leukocytes, bone marrow analysis requires examining up to nearly 40 types of leukocytes. The characteristics of these multiple cell types are often very similar, making fine-grained classification challenging. Traditional machine learning methods and CNN-based leukocyte classification methods have significant limitations. Traditional methods involve cumbersome processes, and the extracted features are often vague, making it difficult to accurately distinguish similar bone marrow cell images. CNN-based methods, on the other hand, suffer from the limitation of the receptive field in convolutional algorithms, which prevents them from learning long-distance dependencies in images. This limitation hinders the accurate extraction of cell image features, thus failing to categorize bone marrow cells with slight differences effectively. To address these challenges, we introduce the SCKansformer model, which leverages Kansformer backbone and advanced attention mechanism to capture intricate patterns and dependencies in cell images, further improving the classification of highly fine-grained datasets.

\section{Methodology}

\subsection{Overall Architecture}

\begin{figure*}
\centering
\includegraphics[width=1.00\textwidth]{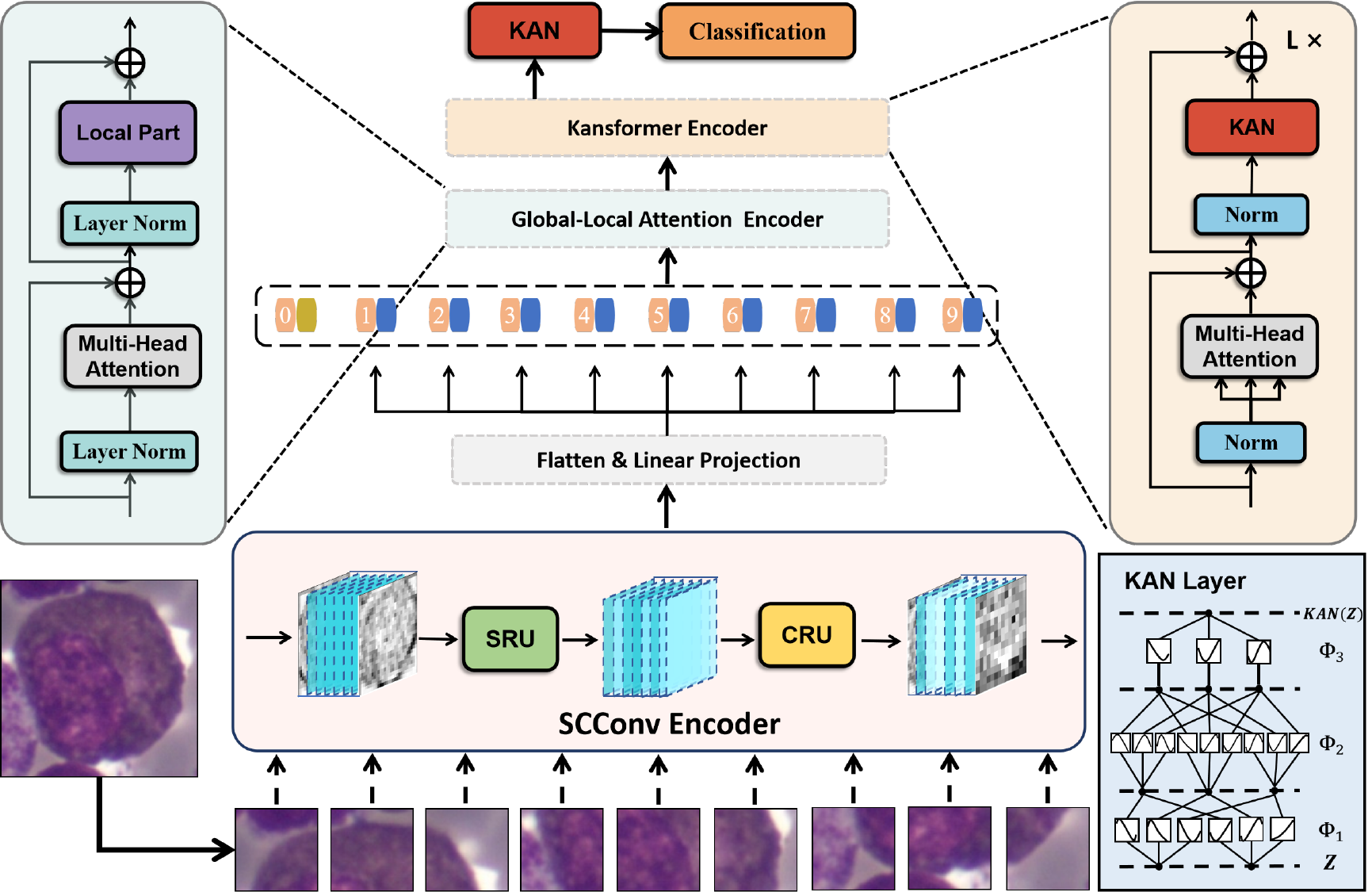}
\caption{The overall architecture of our proposed SCKansformer model. The SCKansformer model primarily comprises three parts: Kansformer Encoder, SCConv Encoder and Global-Local Attention Encoder.} \label{fig1}
\end{figure*}

The structure of the SCKansformer model is illustrated in Fig. \ref{fig1}. The model comprises three main components: the Kansformer Encoder, the SCConv Encoder, and the Global-Local Attention Encoder (GLAE). The Kansformer Encoder enhances the efficiency and interpretability of the Vision Transformer (ViT) model \cite{c25dosovitskiy2020image} by replacing the traditional MLP layer \cite{c36haykin1998neural} with a Kolmogorov-Arnold Network (KAN) \cite{c16liu2024kan}. This substitution improves the nonlinear feature representation and interpretability of the ViT model, with the structure consisting of alternating Multi-head Self-Attention (MSA) layers and KAN networks. The SCConv Encoder comprises a Spatial Reconstruction Unit (SRU) and a Channel Reconstruction Unit (CRU). These units are designed to reduce redundant information in the features extracted from CNNs, specifically targeting spatial and channel redundancy through separation-reconstruction and separation-transformation fusion strategies, respectively. The GLAE consists of an MSA module and a Local Part module for capturing long-range dependencies and local features. The MSA module learns long-range dependencies through the self-attention mechanism, while the Local Part module extracts local features between neighboring pixels using depth-separable convolution. Through the synergistic work of these three components, the SCKansformer model effectively processes microscopic cell images, thereby enhancing both the performance and interpretability of the model. This makes accurate classification of bone marrow blood cell images a possibility.

\subsection{Kansformer Encoder}

MLP layers, also known as fully connected feedforward neural networks, are foundational building blocks in deep learning models and are commonly used to approximate nonlinear functions in machine learning. A MLP comprising $K$ layers can be described as an interplay of transformation matrices ${W}$ and activation functions $\sigma$. This can be mathematically expressed as: 

\begin{equation}
\operatorname{MLP}(\mathbf{Z}) = \left({W}_{K-1} \circ \sigma \circ {W}_{K-2} \circ \sigma \circ \cdots \circ {W}_1 \circ \sigma \circ {W}_0\right) \mathbf{Z}, 
\end{equation}
where it strives to mimic complex functional mappings through a sequence of nonlinear transformations over multiple layers. Despite their widespread use, MLPs have significant drawbacks. In the task of classifying microscopic medical images, specifically, bone marrow blood cell micrographs, the presence of many cell categories with similar features complicates the classification task. MLPs suffer from high computational complexity, high memory consumption, and poor interpretability when processing high-dimensional image data, which can result in inadequate feature learning for different cell categories.

Recently, Liu et al. \cite{c16liu2024kan} proposed the KANs as a promising alternative to MLPs. While MLP are inspired by the generalized approximation theorem, KAN is based on the Kolmogorov-Arnold representation theorem. KANs have a fully connected structure similar to MLPs, but unlike MLPs, which place fixed activation functions on nodes (neurons), KANs use learnable activation functions on edges (weights). Similar to an MLP, a $K$-layer KAN can be characterized as a nesting of multiple KAN layers: 

\begin{equation}
    \operatorname{KAN}(\mathbf{Z})=\left(\boldsymbol{\Phi}_{K-1} \circ \boldsymbol{\Phi}_{K-2} \circ \cdots \circ \boldsymbol{\Phi}_{1} \circ \boldsymbol{\Phi}_{0}\right) \mathbf{Z}, 
\end{equation}
where $\boldsymbol{\Phi}_i$ signifies the $i$-th layer of the entire KAN network. Each KAN layer, with $n_{in}$ -dimensional input and $n_{out}$ -dimensional output,  $\boldsymbol{\Phi}$ comprises $ n_{in} \times n_{out}$ learnable activation functions $\phi$, as shown in Equation (\ref{boldsymbol}): 

\begin{equation}\label{boldsymbol}
    \boldsymbol{\Phi}=\left\{\phi_{k, q, p}\right\}, \quad p=1,2, \cdots, n_{\text {in }}, \quad q=1,2 \cdots, n_{\text {out }}. 
\end{equation}
The pre-activation of $\phi_{k,q,p}$ is simply $Z_{k,p}$; the post-activation of $\phi_{k,q,p}$ is denoted by $\tilde{Z}_{k,q,p}\equiv \phi_{k,q,p}(Z_{k,p})$. The activation value of the $(k+1,q)$ neuron is simply the sum of all incoming post-activations, as shown in Equation(\ref{Z_k+1}): 


\begin{equation}\label{Z_k+1}
    Z_{k+1,q} =  \sum_{p=1}^{n_k} \tilde{Z}_{k,q,p} = \sum_{p=1}^{n_k}\phi_{k,q,p}(Z_{k,p}), \qquad q=1,\cdots,n_{out}.
\end{equation}

Building on these findings, we propose the Kansformer Encoder architecture, which replaces the MLP layers in the Transformer Encoder with KAN layers from the original ViT model to address the aforementioned issues. The structure of the Kansformer Encoder is shown in Fig. \ref{fig1}. It consists of alternating MSA layers and KAN networks. Layer normalization is applied before each block, and residual concatenation is used after each block. In this architecture, a microscopic image \( X \in \mathbb{R}^{H \times W \times C} \) is reshaped into a series of 2D image block sequences \( X_p \in \mathbb{R}^{N \times (P^2 \cdot C)} \) after a Patch Embedding operation, where \( (H, W) \) is the resolution of the original image, \( C \) is the number of channels, \( (P, P) \) is the resolution of each image block, and \( N = \frac{HW}{P^2} \) is the number of blocks obtained. Subsequently, the SCConv Encoder processes the sequence to address spatial and channel redundancy features, and the processed result is used as the effective input sequence for the Kansformer.

The Kansformer uses a constant hidden vector size \( D \) across all its layers, so image blocks are flattened and mapped to \( D \) dimensions using a trainable linear projection, known as Patch Embedding. Similar to BERT's class token, an embedding is added before the embedded patch sequence, and its state at the output of the Kansformer Encoder serves as the image classification flag. After assembling the input embedding vectors of class tokens, image block embeddings, and positional encodings into one sequence, the GLAE captures relationships between neighboring pixels. The processed feature vectors are then fed into the Kansformer Encoder. The Kansformer Encoder is composed of serially stacked Kansformer Encoder Blocks. Finally, the features corresponding to the learned class token are extracted for image classification. This architecture effectively addresses the limitations of traditional MLPs and enhances the performance and interpretability of microscopic medical image classification models.

\subsection{Global-Local Attention Encoder}

The variations in color and quality of microscopic images often arise due to differences in equipment and imaging conditions. Furthermore, the long-tailed distribution and low resolution inherent in the myeloid blood cell dataset pose significant challenges, limiting the model's ability to recognize and classify rare cell classes. As depicted in Fig.~\ref{fig1}, the core design of the GLAE integrates the MSA module to capture long-range dependencies and the Local Part module to extract local features from neighboring pixels. These components are harmonized through an adaptive weight-sharing mechanism, ensuring that both global and local features are processed seamlessly throughout the network. This balanced approach enables the model to address the dataset's complexities by capturing extensive dependencies while effectively extracting detailed local features using depth-separable convolution. This dual approach allows the model to capture complex relationships between neighboring pixels, enhancing its ability to discern subtle differences in cell morphology.

\subsubsection{MSA Module}
This module calculates different weights for different objects by learning from the interactions between input vectors (self-attention mechanism) to identify regions of greater importance. The self-attention calculation formulas are shown in Equations (\ref{eqcosqk}) and (\ref{eqatt}): 

\begin{equation}\label{eqcosqk}
    \text{Cos}(Q,K) = \text{Softmax}\left(\frac{QK^T}{\sqrt{d}}\right),
\end{equation}
\begin{equation}\label{eqatt}
    Attention(Q,K,V) = \text{Cos}(Q,K) \times V.
\end{equation}
Self-attention is so termed because the matrices \( Q \), \( K \), and \( V \) are all derived from a sequence of input vectors \( X \in \mathbb{R}^{N \times d} \) via linear mappings, as shown in Equation (\ref{eqQKV}): 

\begin{equation}\label{eqQKV}
\begin{aligned}
    Q &= XW_Q, \\
    K &= XW_K, \\
    V &= XW_V,
\end{aligned}
\end{equation}
where the weight matrices \( W_Q, W_K, \) and \( W_V \) are elements of \( \mathbb{R}^{d \times d} \) and \( Q \) denotes the query information, \( K \) denotes the key information (used to match \( Q \)), and \( V \) represents the content vector. As shown in Equations (\ref{eqcosqk}) and (\ref{eqatt}), the inner product of \( Q \) and \( K \) (i.e., \( QK^T \)) captures the interrelated relationships between feature points. This product is then scaled by \( \sqrt{d} \) where \( d \) is the dimension of the hidden layer. This scaling serves two purposes: 1). To prevent the values from becoming too large when subsequently using the Softmax function, which could result in partial derivatives tending toward zero.
2). To normalize \( QK^T \), ensuring it has an expected distribution with a mean of 0 and a variance of 1.

The Softmax function is then applied to obtain the weight information of each object, as shown in Equation (\ref{eqsoftzi}): 

\begin{equation}\label{eqsoftzi}
    \text{softmax}(z_i) = \frac{e^{z_i}}{\sum_{c=1}^C e^{z_c}},
\end{equation}
where \( e \) denotes the natural constant, \( z_i \) denotes the \( i \)-th element of the vector \( z \), and \( \sum_{c=1}^C e^{z_c} \) represents the sum of the exponentials of all elements in the vector \( z \). The purpose of the Softmax function is to generate probability distributions for the different elements in \( z \), thereby obtaining the weight information for each object in the vector \( QK^T/\sqrt{d} \). This weight information is finally used to obtain a weighted output vector by matching it with \( V \).

To facilitate parallel attention to input image sequences from multiple perspectives and to enhance the model's expressiveness and generalization capabilities, the MSA mechanism is employed. The MSA mechanism performs multiple attention computations for each input image sequence, focusing on its relationships with other sequences from various perspectives. This enables the model to collectively focus on information from different representational subspaces at different locations, capturing more comprehensive and fine-grained global semantic information. The computational formula for the MSA mechanism involves concatenating different attention heads in parallel operations, which improves the model's efficiency and speeds up the training and inference processes, as shown in Equation (\ref{eqmha}): 

\begin{multline}\label{eqmha}
    \text{MHA}(Q,K,V) = \text{Concat}(Attention(Q_1,K_1,V_1), \ldots,\\Attention(Q_i,K_i,V_i)).
\end{multline}

\subsubsection{Local Part Module}
\begin{figure*}
\centering
\includegraphics[width=1.00\textwidth]{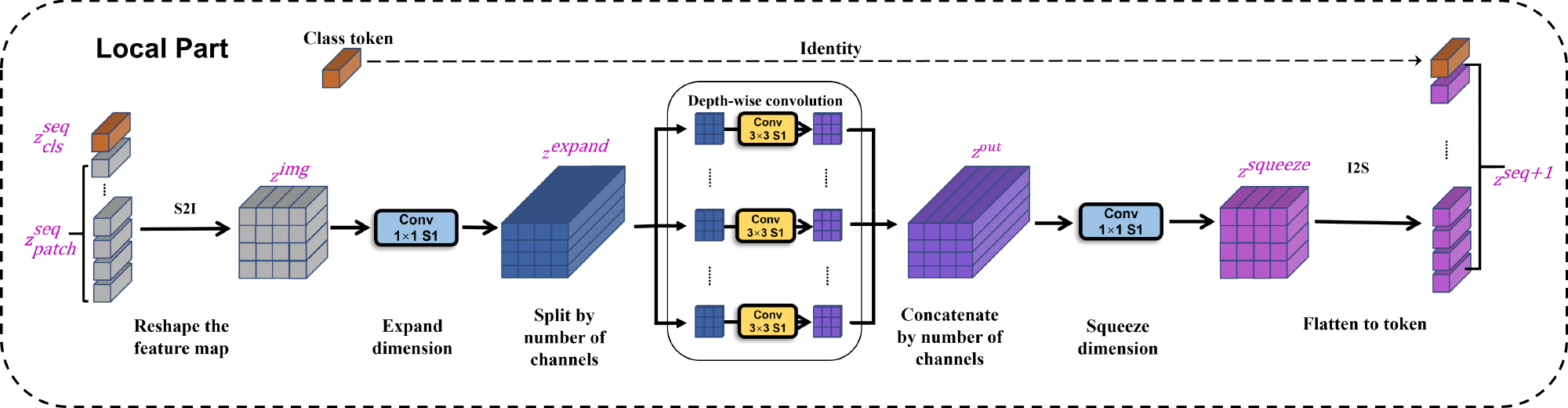}
\caption{The framework of Global-Local Attention Encoder. The Global-Local Attention Encoder combines the MSA module and the Local Part module to effectively capture the global and local features of microscopic images, which enhances the model's ability to recognize long-distance dependencies and fine-grained features.} \label{fig2}
\end{figure*}

Due to the exclusive use of the self-attention mechanism for spread feature maps, only global dependencies between all input image sequences are captured. This can result in a lack of information interaction between neighboring pixels, i.e., local dependencies between nearby pixels are not effectively modeled. To address this issue, we propose a localization module that models local dependencies between neighboring pixels by introducing depth-separable convolution. The localization module adopts a convolutional approach to learn local features. This approach first expands the dimensionality and then restores it to the original dimensionality, enabling the learning of more complex nonlinear relationships between features. The main flow of the Localization Module is illustrated in Fig. \ref{fig2}:

1). The serialized vectors are initially divided into a trainable serialized vector \( z_{cls}^{seq} \) and an image serialized vector \( z_{patch}^{seq} \).
2). The image serialized vectors are then reconstructed into a two-dimensional feature map through the Sequence to Image (S2I) operation, which aims to recover the physical location and adjacency between pixels.
3). A \( 1 \times 1 \) convolution kernel with a stride of 1 is used to expand the dimensionality, followed by BN and the h-swish activation function (as shown in Equation (\ref{eqh-s})) to address the gradient vanishing problem and add nonlinearity to the model.
4). After expanding the dimensionality, the feature map \( z^{out} \) is obtained using \( 3 \times 3 \) depth-wise convolution with a stride of 1.
5). The dimensionality is then recovered using a \( 1 \times 1 \) convolution with a stride of 1.
6). Upon completing the local feature extraction, the feature map is re-expanded into a one-dimensional serialized vector \( z_{patch}^{out} \).
7). Finally, the one-dimensional serialized vector is concatenated with the trainable serialized vector, serving as the input vector for the next layer. The process can be represented by the following series of equations: 

\begin{equation}\label{eqh-s}
    h - swish(x) = x * sigmoid(x),
\end{equation}
\begin{equation}\label{eqzc-zp}
    z^{seq}_{cls},z^{seq}_{patch} = Split(z^{seq}),
\end{equation}
\begin{equation}\label{eqzimg}
    z^{img} = S2I(z^{seq}_{patch}),
\end{equation}
\begin{equation}\label{eqzexpand}
    z^{expand} = h - swish(BN(Conv(z^{img}))),
\end{equation}
\begin{equation}\label{eqzout}
    z^{out} = DW(z^{expand}),
\end{equation}
\begin{equation}\label{eqzsqueeze}
    z^{squeeze} = BN(Conv(z^{out})),
\end{equation}
\begin{equation}\label{eqzop}
    z^{out}_{patch} = I2S(z^{squeeze}),
\end{equation}
\begin{equation}\label{eqs+1}
    z^{seq+1} = Concat(z^{seq}_{cls},z^{out}_{patch}).
\end{equation}

\subsection{SCConv Encoder}

Microscopic images of bone marrow blood cells often contain complex backgrounds and cells in various states, leading to the extraction of convoluted and confusing features that include a large amount of redundant information. This redundant information is typically unrelated to the target cell classification and only serves to increase the computational complexity and potential error of the model. To address this issue, the SCConv Encoder module is employed to remove redundant information and simplify computational demands by deeply reconstructing spatial and channel information \cite{c35li2023scconv}.

\begin{figure*}
\centering
\includegraphics[width=1.00\textwidth]{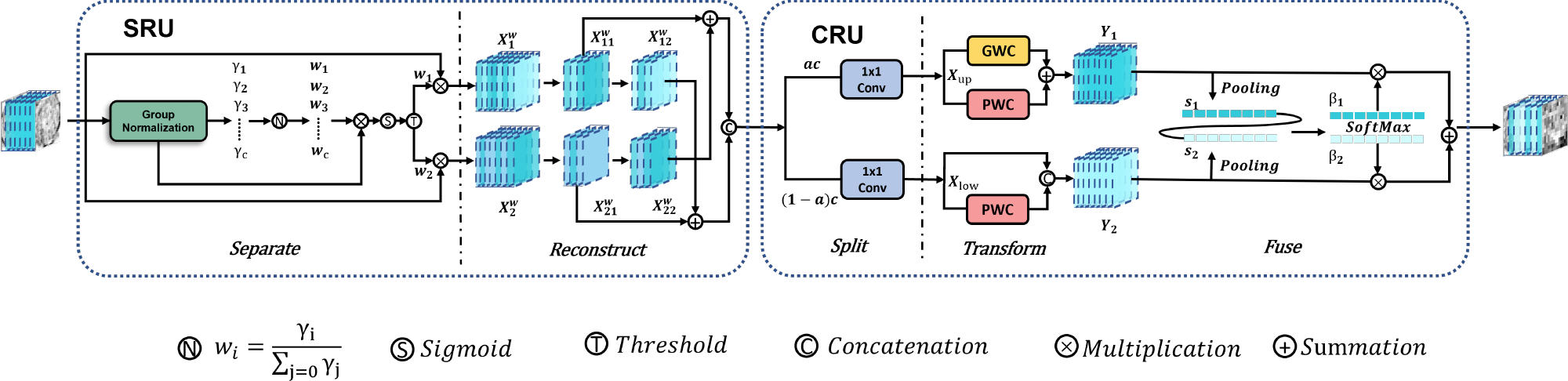}
\caption{The framework of SCConv Encoder. The SCConv Encoder optimizes the feature representation by means of Spatial Reconstruction Units and Channel Reconstruction Units, which reduces feature redundant information of microscopic images.} \label{fig3}
\end{figure*}

As illustrated in Fig. \ref{fig3}, the SCConv Encoder module comprises two primary units: the SRU and the CRU. The SRU reduces spatial redundancy through a separation-reconstruction method, whereas the CRU reduces channel redundancy via a segmentation-transformation-fusion method. Together, these two units effectively reduce the redundant information in features extracted by the CNN.

\subsubsection{SRU Module} 
The input microscopic images contain miscellaneous sequence information following the Patch Embedding operation. To utilize the spatial redundancy of the features effectively, we perform the spatial refinement operation using the SRU module. The SRU module reduces spatial redundancy through a separation-reconstruction strategy.

Initially, the input feature maps are processed using Group Normalization (GN) to distinguish between information-rich feature maps and less informative ones, as shown in Equation (\ref{eqxout}): 

\begin{equation}\label{eqxout}
    X_{\text{out}} = GN(X) = \gamma  \frac{X - \mu}{\sqrt{\sigma^2 + \epsilon}}  + \beta,
\end{equation}
where the input feature map \( X \in \mathbb{R}^{N \times C \times W \times H} \), \( N \) represents the batch axis, \( C \) the channel axis, and \( H \) and \( W \) the spatial height and width axes, respectively. In this context, \( \mu \) and \( \sigma \) denote the mean and standard deviation within \( X \), \( \varepsilon \) is a small positive constant added to ensure numerical stability during division, and \( \gamma \) and \( \beta \) are trainable affine transformation parameters. The scaling factor \(\gamma\) subsequently evaluates the amount of information carried by these different feature maps, as shown in Equation (\ref{eqw_gamma}): 

\begin{equation}\label{eqw_gamma}
    W_\gamma = \{w_i\} = \frac{\gamma_i}{\sum_{j=1}^{C} \gamma_j}, \quad i,j=1,2,\ldots,C.
\end{equation}
The normalized correlation weight \(W_\gamma\) is obtained using the scaling factor \(\gamma\) as a metric, which measures the importance of the various feature maps.

The obtained correlation weights \(W_\gamma\) are then mapped to the range of 0 to 1 using a sigmoid function, as shown in Equation (\ref{eqXr}): 

\begin{equation}\label{eqXr}
    X_\gamma = \text{Gate}(\text{Sigmoid}(W_\gamma (\text{GN}(X)))).
\end{equation}
These weights are then gated by a threshold (set to 0.5 in the experiment). Weights exceeding the threshold are assigned a value of 1 to generate the informative weight \(W_1\), while weights below the threshold are assigned a value of 0 to obtain the non-informative weight \(W_2\). Subsequently, the input feature \(X\) is multiplied with the two weighted features separately to derive the informative key feature \(X_1^w\) and the less informative redundant feature \(X_2^w\).

Finally, a cross-reconfiguration operation is performed to fully merge the two weighted informative features, enhancing the information flow between them and ultimately generating the spatially refined feature map \(X^w\).

\subsubsection{CRU Module}

Given that spatial redundancy may still persist in spatial refinement, the CRU module is introduced to enhance performance. This module employs a separation-transformation-fusion strategy to reduce feature redundancy by replacing standard convolution with three operations: Split, Transform, and Fuse.

The Split operation divides the input spatially refined feature \(X_w\) into two parts: one with \(\alpha C\) channels and the other with \((1-\alpha)C\) channels, where \(\alpha\) is a hyperparameter set experimentally to 0.5. The channel numbers of these two feature sets are then compressed using a \(1 \times 1\) convolution kernel to obtain \(X_{up}\) and \(X_{low}\), respectively.

The Transform operation processes \(X_{up}\) using the 'Rich Feature Extractor', which performs Group-wise Convolution (GWC) and Point-wise Convolution (PWC). GWC reduces the number of parameters and computational effort, while PWC compensates for information loss and facilitates information flow across feature channels. The combined result of these operations is added to produce the output \(Y_1\). Concurrently, PWC is applied to \(X_{low}\) as a complementary operation to the Rich Feature Extractor, and the result is concatenated with the original input to yield \(Y_2\).

The Fuse operation employs a streamlined SKNet methodology \cite{c40li2019selective} to adaptively integrate \(Y_1\) and \(Y_2\). Initially, this procedure amalgamates global spatial information with channel statistics via global average pooling, yielding the pooled outcomes \(S_1\) and \(S_2\). Subsequently, a Softmax operation is applied to \(S_1\) and \(S_2\) to obtain the feature weight vectors \(\beta_1\) and \(\beta_2\). Finally, the outputs are calculated using these feature weight vectors, resulting in \(Y = \beta_1 Y_1 + \beta_2 Y_2\), where \(Y\) represents the feature post channel refinement.

\begin{figure*}
\centering
\includegraphics[width=1.00\textwidth]{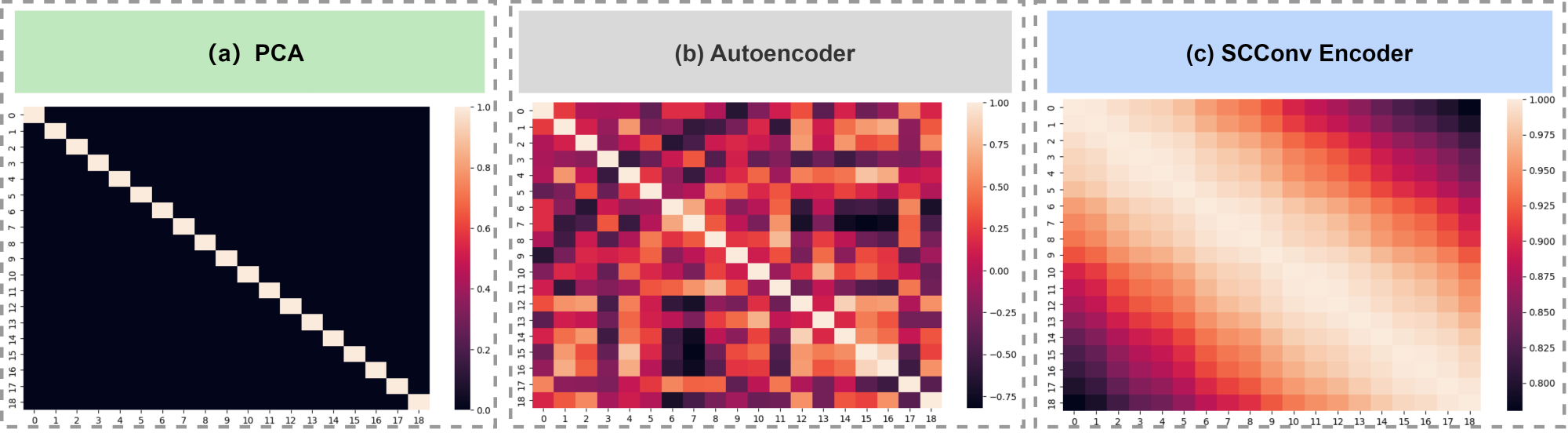}
\caption{The feature dimensionality reduction visualizations of SCConv Encoder and other traditional methods. (\textbf{a}) Heatmap of PCA. (\textbf{b})  Heatmap of Autoencoder. (\textbf{c}) Heatmap of SCConv Encoder. } \label{fig4}
\end{figure*}

The primary advantage of the SCConv Encoder is its ability to target specific spatial and channel redundancies in microscopic images for feature compression. Unlike traditional dimensionality reduction techniques that mainly rely on statistical variance for feature reduction, the SCConv Encoder excels in preserving the spatial structural information crucial for accurate classification. Traditional methods such as PCA and Autoencoders, while effective, often struggle to maintain this essential information. In contrast, the SCConv Encoder significantly enhances classification accuracy by effectively reducing redundant information, particularly in high-dimensional microscopic image data. These benefits are further illustrated through visualizations, as shown in Fig.~\ref{fig4}.

\section{Experiment}
\subsection{Data Acquisition and Processing}

Due to the difficulties associated with collecting and labeling images of bone marrow blood samples, and the highly granular classification required for bone marrow blood cells, there is a notable lack of accurately labeled, large-scale, fine-grained bone marrow blood cell datasets in the academic community. To address this gap, our team, in collaboration with the Department of Hematology at Zhejiang Hospital in Hangzhou, Zhejiang Province, China, has established the private BMCD-FGCD dataset. This research study was conducted retrospectively and in accordance with the principles of the Declaration of Helsinki. We applied for an informed consent waiver. Approval was granted by the Ethics Review Committee of Zhejiang Hospital: 2024 Proc. No. (023K). 
 
\begin{figure*}
\centering
\includegraphics[width=1.00\textwidth]{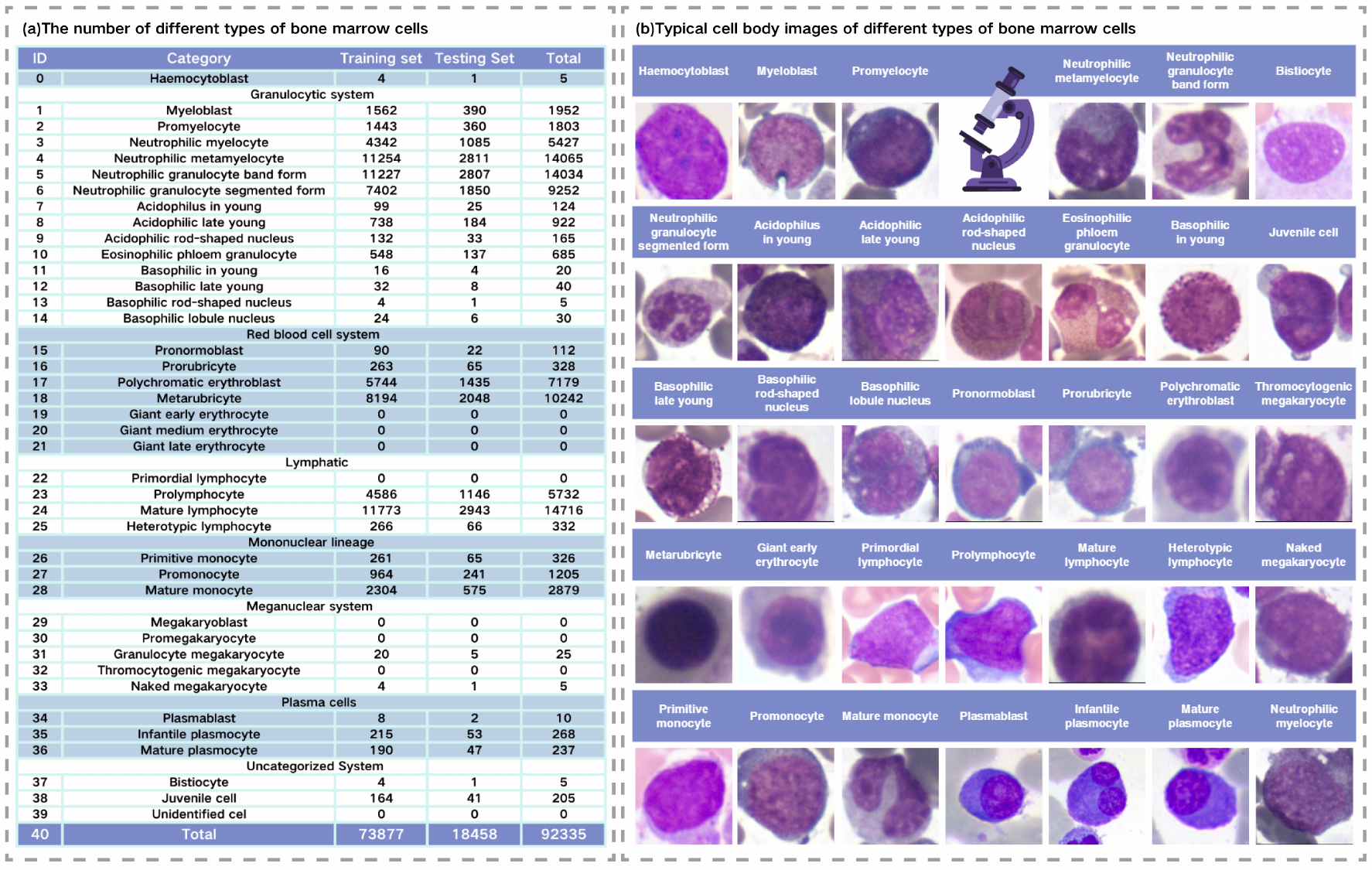}
\caption{(\textbf{a}) Statistical presentation on the number of different types of bone marrow blood cells in our BMCD-FGCD dataset. Our dataset is divided according to categories, the training set contains 73877 samples and the testing set contains 18458 samples, and the ratio of the training set to the testing set is 8:2. (\textbf{b}) Typical cell body images of different types of bone marrow blood cells in our BMCD-FGCD dataset. Intuitive visualization of information such as morphological features, color and structure of various types of bone marrow blood cells.} \label{fig5}
\end{figure*}

This dataset, which contains over 10,000 images, comprises nearly 40 different classifications and has been accurately labeled by three senior physicians with more than 20 years of experience in hematology. The BMCD-FGCD dataset includes a total of 92,335 bone marrow blood cell images that span nearly 40 distinct blood cell categories. For the purposes of this study, the dataset is divided into a training set containing 73,877 samples and a test set containing 18,458 samples. Detailed statistical information is presented in Fig. \ref{fig5}. In an effort to contribute positively to the advancement of the field, we have decided to make the BMCD-FGCD dataset publicly available to other researchers.

Our private dataset comprises original images derived from bone marrow sample images within the hospital data system. These images have been collected from patients who presented to the Department of Hematology at Zhejiang Hospital and were deemed by specialized physicians to require bone marrow aspiration for diagnostic assistance following preliminary clinical evaluations. The construction of the BMCD-FGCD dataset, as illustrated in Fig. \ref{fig6}, was methodically conducted through a tripartite process:

\begin{figure*}
\centering
\includegraphics[width=1.00\textwidth]{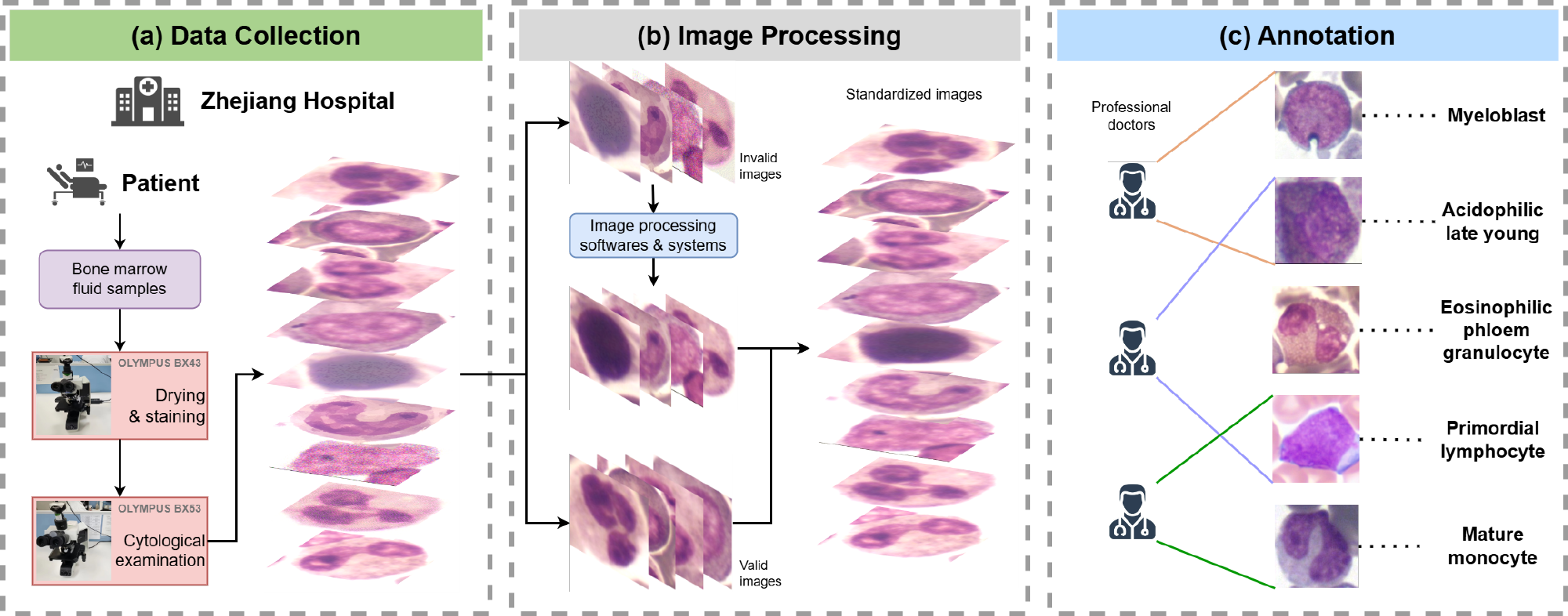}
\caption{Workflow of the establishment of the proposed dataset. (\textbf{a}) Data were extracted from patients' bone marrow fluid samples obtained at the Department of Hematology at Zhejiang Hospital. The preparation and examination of the smears were carried out using two types of Olympus microscopes. (\textbf{b}) Images deemed to be of low quality underwent standardization through image processing software and systems to ensure their suitability for algorithmic processing. (\textbf{c}) Expert physicians at Zhejiang Hospital meticulously identified and manually annotated each image with the respective cell type.} \label{fig6}
\end{figure*}

\subsubsection{Acquisition Stage}
Patients are positioned either prone or laterally under the careful supervision of physicians. The skin overlying the puncture site is thoroughly disinfected, and a local anesthetic is administered to achieve targeted infiltration. Utilizing a sterile, disposable puncture device equipped with a micropuncture needle, the clinician advances the needle through the cortical bone and into the marrow cavity. Upon successful entry into the marrow cavity, the physician withdraws the needle's core and attaches a 10 ml syringe to aspirate the bone marrow fluid, which is subsequently prepared as a smear sample. After being dried and stained, the smear samples are examined cytologically under a microscope by a specialized physician. We selected high-quality images of patients of different ages, genders, and diseases from Zhejiang Hospital's electronic medical record system.

\subsubsection{Image Processing Stage}
Acquired images of bone marrow blood cells were imported into specialized image processing software or systems to ensure they conformed to the requisite standards for subsequent analysis and annotation. For images that did not meet these specified criteria, pre-processing techniques, such as clarity enhancement or normalization procedures, were employed to ensure the quality of the images was suitable for further algorithmic processing.

\subsubsection{Labeling Stage}
This phase entails the meticulous identification and categorization of each bone marrow blood cell subtype, encompassing lymphocytes, neutrophils, monocytes, among others. The classification process was underpinned by medical expertise with exceptional detail and accuracy, in order to guarantee both the precision and the integrity of the image dataset. During this phase, three senior physicians on the team with more than 20 years of experience in hematology classified and annotated the nearly 40 different bone marrow blood cells in this dataset. Subsequently, the team processed the categorized data with another high degree of precision and accuracy to ensure the quality of the microscopic images.

\subsection{Comparison to Existing Datasets}

In order to validate the generalization of the model, we also used two publicly available leukocyte datasets for further validation, namely the PBC dataset and the ALL-IDB dataset. Since our private dataset is a category-unbalanced myeloid cell dataset, when choosing the public leukocyte dataset for validation, the PBC and ALL-IDB datasets we chose were category-balanced.

\textbf{ALL-IDB dataset} \cite{c31labati2011all}: This dataset is a specialized resource developed for the segmentation and classification of mother cells (immature or precursor cells) in Acute Lymphocytic Leukemia (ALL). The images in the dataset are captured using a Canon PowerShot G5 camera, saved in JPG format with a 24-bit color depth, and the dataset is divided into two subsets: ALL-IDB1, which is dedicated to segmentation, and ALL-IDB2, which is tailored for classification tasks. ALL-IDB1 comprises 108 labeled peripheral blood smear images with a resolution of 2,592 × 1,944 pixels, featuring approximately 39,000 blood cells, including 510 lymphocytes from ALL patients validated by expert oncologists. ALL-IDB2 contains images cropped from the ALL-IDB1 collection that display acute lymphoblasts, each standardized to a resolution of 257 × 257 pixels. This subset includes 260 images, of which half portray lymphoblasts from ALL patients.

The ALL-IDB dataset has a narrow scope of application, being suitable only for segmenting and classifying ALL mother cells. It cannot be applied effectively to the broader and more critical task of classifying bone marrow leukocytes. Additionally, the ALL-IDB2 classification dataset contains only 260 peripheral blood cell smear images, which is insufficient for large model training. Moreover, it only includes peripheral blood images and does not cover bone marrow blood smears, which are more important for diagnosing leukemia.

\textbf{PBC dataset} \cite{c32acevedo2020dataset}: This dataset comprises a collection of blood cell images derived from peripheral blood samples of healthy volunteers. It features a compendium of 17,092 images of individual normal cells captured using a CellaVision DM96 analyzer, located in the core laboratory of the Hospital Clinic of Barcelona. The dataset is meticulously categorized into eight cell categories: neutrophils, eosinophils, basophils, lymphocytes, monocytes, immature granulocytes (including promyelocytes, myelocytes, and metamyelocytes), erythrocytes, and platelets or thrombocytes. The images, standardized at a resolution of 360 × 363 pixels and in JPG format, are classified by expert clinical pathologists. The source of these images is individuals who are not affected by infections, hematologic malignancies, oncologic diseases, and who are not undergoing any form of medication treatment.

The PBC dataset similarly only includes blood images from peripheral blood smears and does not cover the more critical bone marrow sections. Although the dataset is refined into eight blood cell types compared to the ALL-IDB dataset, it still focuses on peripheral blood smears, which do not meet the requirements for a classification model for bone marrow sections, especially when our dataset includes up to forty cell classifications for bone marrow sections.


\subsection{Experimental Settings}

The method was implemented in Python using the PyTorch deep learning framework and the PyCharm Integrated Development Environment (IDE). Experiments were conducted on a single GPU (NVIDIA Tesla V100) running Ubuntu 18.04. Pre-training was performed on the ImageNet-1K dataset \cite{c39deng2009imagenet}. The training process lasted for 100 epochs with a batch size of 128, taking approximately 12.6 hours.

For the training data, input images were randomly cropped to $224 \times 224$ pixels, and data augmentation techniques, such as random horizontal flipping, were applied to enhance the dataset. The Adam optimizer was used in conjunction with a cosine annealing algorithm. At the same time, for the testing data, images were resized to $256 \times 256$ pixels and then center-cropped to $224 \times 224$ pixels.

\subsection{Evaluation Indicators}

To comprehensively evaluate the performance of different models on the category-unbalanced BMCD-FGCD dataset, we use the following standard evaluation metrics: Accuracy, Precision, Recall, F1-score, Balanced Accuracy, and MCC. These metrics are calculated using the formulas shown below: 

\begin{equation}
    \text{Precision} = \frac{TP}{TP + FP},
\end{equation}
\begin{equation}
    \text{Recall} = \frac{TP}{TP + FN},
\end{equation}
\begin{equation}
    \text{F1-score} = 2 \cdot \frac{\text{Precision} \cdot \text{Recall}}{\text{Precision} + \text{Recall}},
\end{equation}
\begin{equation}
    \text{Accuracy} = \frac{TP + TN}{TP + FP + TN + FN},
\end{equation}
\begin{equation}
    \text{Balanced Accuracy} = \frac{1}{2} ( \frac{TP}{TP + FN} + \frac{TN}{TN + FP} ),
\end{equation}
\begin{equation}
    \text{MCC} = \frac{ TP \times TN - FP \times FN }{\sqrt{(TP + FP)(TP + FN)(TN + FP)(TN + FN)}},
\end{equation}
where True Positive (TP), False Positive (FP), False Negative (FN), and True Negative (TN) are defined as follows: TP and TN denote samples correctly predicted as true and false, respectively, while FP and FN denote samples incorrectly predicted as true and false, respectively.

\subsection{Comparison Experiment}
\subsubsection{Comparison of Model Performance}
As shown in Table \ref{tab1}, we further validate our model by experimentally comparing it with other typical classification models and advanced microscopic image cell classification models on the BMCD-FGCD dataset. Compared with other typical or state-of-the-art models, SCKansformer exhibits significant advantages in handling the microscopic image classification task.



When compared with advanced classification models such as ViT and the lightweight EfficientNetV2, the advantages of the SCKansformer become even more evident. The SCKansformer model improves Accuracy by 5.81 and 6.03 percentage points compared to ViT and EfficientNetV2, respectively. This significant enhancement is primarily due to the SCKansformer’s SCConv Encoder and Global-Local Attention Mechanism, which effectively capture the fine-grained features of microscopic images. This capability addresses the shortcomings of ViT and EfficientNetV2 in processing high-resolution microscopic images, particularly in terms of local feature capture.

Compared with traditional leukemia cell classification models like WBCsNet and BloodCaps, the SCKansformer model demonstrates superior performance across various metrics. In terms of Accuracy, the SCKansformer model surpasses the aforementioned models by 10.80 and 4.92 percentage points, respectively. By introducing the KAN network and SCConv Encoder, the SCKansformer effectively addresses the limited feature extraction capability and parameter redundancy issues inherent in the traditional CNN model WBCsNet. Additionally, through the incorporation of the GLAE Encoder, the SCKansformer compensates for the high computational complexity and insufficient global information capture associated with BloodCaps’ capsule networks.

At the same time, we compare our SCKansformer model with the advanced microscopic cell classification models WBC-GLAformer and WBC YOLO-ViT, both of these models exhibit limitations in handling feature redundancy and nonlinear feature representation. They are particularly challenged in the task of classifying complex microscopic cell images and have limited effectiveness in capturing subtle features and distinguishing similar cell images. Specifically, WBC-GLAformer and WBC YOLO-ViT achieve F1-scores of 79.60 and 79.67, respectively, while the SCKansformer model attains an F1-score of 84.34, representing improvements of 4.74 and 4.67 percentage points. In terms of Accuracy, WBC-GLAformer and WBC YOLO-ViT both register scores of 79.45 and 78.47, respectively, whereas SCKansformer achieves an Accuracy of 83.23, marking enhancements of 3.78 and 4.76 percentage points.

\begin{table*}[h]
\centering
\caption{Comparison of SCKansformer with other state-of-the-art models on the private BMCD-FGCD dataset.}
\label{tab1}
\begin{tabularx}{\textwidth}{
    >{\centering\arraybackslash\hsize=0.7\hsize}X
    >{\centering\arraybackslash\hsize=0.5\hsize}X
    >{\centering\arraybackslash\hsize=0.5\hsize}X
    >{\centering\arraybackslash\hsize=0.5\hsize}X
    >{\centering\arraybackslash\hsize=0.5\hsize}X
    >{\centering\arraybackslash\hsize=0.5\hsize}X
  }
\toprule
\textbf{Method} & \textbf{Precision} & \textbf{Recall} & \textbf{F1-score} & \textbf{Accuracy} & \textbf{Balanced Accuracy}\\
\hline
VGG16 \cite{c23simonyan2014very} & 74.68 & 74.43 & 74.55 & 74.50 & 35.79 \\
VGG19 \cite{c23simonyan2014very} & 73.53 & 73.35 & 73.44 & 73.26 & 37.63 \\
ResNet-50 \cite{c24he2016deep} & 75.62 & 75.49 & 75.55 & 75.48 & 49.76 \\
ResNet-101 \cite{c24he2016deep} & 75.43 & 75.20 & 75.31 & 75.21 & 50.50 \\
WBCsNet \cite{c27baghel2022wbcs} & 73.65 & 72.04 & 72.84 & 72.43 & 47.50 \\
BloodCaps \cite{c28long2021bloodcaps} & 78.72 & 76.74 & 77.94 & 78.31 & 54.31 \\
ViT \cite{c25dosovitskiy2020image} & 77.65 & 77.41 & 77.53 & 77.42 & 70.47 \\
EfficientV2 \cite{c26tan2021efficientnetv2} & 77.37 & 77.21 & 77.29 & 77.20 & 71.51 \\
WBC-GLAformer \cite{c30chen2023fine} & 79.82 & 79.38 & 79.60 & 79.45 & 74.31 \\
WBC YOLO-ViT \cite{c29tarimo2024wbc} & 78.92 & 79.31 & 79.67 & 78.47 & 77.41 \\
\textbf{SCKansformer (Ours)} & \textbf{85.82} & \textbf{84.28} & \textbf{84.34} & \textbf{83.23} & \textbf{82.91} \\
\bottomrule
\end{tabularx}
\end{table*}

\subsubsection{Validation of Model Generalizability}
To further validate the generalizability of our proposed SCKansformer model, we compare it with other typical classification models and advanced microimaging cell classification models on two publicly available leukocyte datasets, PBC and ALL-IDB. As shown in Table \ref{tab2} and Table \ref{tab3}, the SCKansformer model exhibits strong generalizability and achieves superior results on both datasets.
The generalizability validation further demonstrates that our proposed SCKansformer model performs well on cellular imaging datasets with uneven categories and adapts effectively to datasets with balanced categories, such as ALL-IDB and PBC, yielding promising results.

\begin{table*}[h]
\centering
\caption{Comparison with other state-of-the-art models on the PBC public dataset.}
\label{tab2}
\begin{tabularx}{\textwidth}{
    >{\centering\arraybackslash\hsize=0.7\hsize}X
    >{\centering\arraybackslash\hsize=0.5\hsize}X
    >{\centering\arraybackslash\hsize=0.5\hsize}X
    >{\centering\arraybackslash\hsize=0.5\hsize}X
    >{\centering\arraybackslash\hsize=0.5\hsize}X
    >{\centering\arraybackslash\hsize=0.5\hsize}X
  }
\toprule
\textbf{Method} & \textbf{Precision} & \textbf{Recall} & \textbf{F1-score} & \textbf{Accuracy} & \textbf{Balanced Accuracy}\\
\hline
VGG16 \cite{c23simonyan2014very} & 98.02 & 98.01 & 98.01 & 98.01 & 98.30 \\
VGG19 \cite{c23simonyan2014very} & 97.31 & 97.28 & 97.29 & 97.28 & 97.25 \\
ResNet-50 \cite{c24he2016deep} & 98.83 & 98.83 & 98.83 & 98.83 & 98.99 \\
ResNet-101 \cite{c24he2016deep} & 98.83 & 98.83 & 98.83 & 98.83 & 98.99 \\
WBCsNet \cite{c27baghel2022wbcs} & 97.34 & 97.33 & 97.33 & 97.33 & 96.93 \\
BloodCaps \cite{c28long2021bloodcaps} & 98.80 & 97.77 & 97.77 & 97.77 & 97.25 \\
ViT \cite{c25dosovitskiy2020image} & 96.90 & 96.90 & 96.89 & 96.90 & 96.75 \\
EfficientV2 \cite{c26tan2021efficientnetv2} & 98.74 & 98.74 & 98.74 & 98.74 & 98.84 \\
WBC-GLAformer \cite{c30chen2023fine} & 99.40 & 99.39 & 99.39 & 99.40 & 98.79 \\
WBC YOLO-ViT \cite{c29tarimo2024wbc} & 98.83 & 98.82 & 98.82 & 98.82 & 98.31 \\
\textbf{SCKansformer (Ours)} & \textbf{99.47} & \textbf{99.46} & \textbf{99.46} & \textbf{99.46} & \textbf{99.37} \\
\bottomrule
\end{tabularx}
\end{table*}


\begin{table*}[h]
\centering
\caption{Comparison with other state-of-the-art models on the ALL-IDB public dataset.}
\label{tab3}
\begin{tabularx}{\textwidth}{
    >{\centering\arraybackslash\hsize=0.7\hsize}X
    >{\centering\arraybackslash\hsize=0.5\hsize}X
    >{\centering\arraybackslash\hsize=0.5\hsize}X
    >{\centering\arraybackslash\hsize=0.5\hsize}X
    >{\centering\arraybackslash\hsize=0.5\hsize}X
    >{\centering\arraybackslash\hsize=0.5\hsize}X
  }
\toprule
\textbf{Method} & \textbf{Precision} & \textbf{Recall} & \textbf{F1-score} & \textbf{Accuracy} & \textbf{MCC}\\
\hline
VGG16 \cite{c23simonyan2014very} & 96.43 & 96.15 & 96.15 & 96.15 & 92.58 \\
VGG19 \cite{c23simonyan2014very} & 98.15 & 98.08 & 98.08 & 98.08 & 89.06 \\
ResNet-50 \cite{c24he2016deep} & 96.15 & 96.15 & 96.15 & 96.15 & 92.58 \\
ResNet-101 \cite{c24he2016deep} & 96.15 & 96.15 & 96.15 & 96.15 & 88.53 \\
WBCsNet \cite{c27baghel2022wbcs} & 95.87 & 95.69 & 95.69 & 95.69 & 93.25 \\
BloodCaps \cite{c28long2021bloodcaps} & 97.32 & 97.31 & 97.31 & 97.31 & 96.44 \\
ViT \cite{c25dosovitskiy2020image} & 98.15 & 98.08 & 98.08 & 98.08 & 96.22 \\
EfficientV2 \cite{c26tan2021efficientnetv2} & 98.15 & 98.08 & 98.08 & 98.08 & 96.23 \\
WBC-GLAformer \cite{c30chen2023fine} & 99.15 & 99.10 & 99.09 & 99.10 & 97.52 \\
WBC YOLO-ViT \cite{c29tarimo2024wbc} & 97.32 & 97.27 & 97.27 & 97.28 & 96.21 \\
\textbf{SCKansformer (Ours)} & \textbf{99.90} & \textbf{99.85} & \textbf{99.85} & \textbf{99.86} & \textbf{99.83} \\
\bottomrule
\end{tabularx}
\end{table*}


\subsection{Ablation Experiment}
\subsubsection{Ablation of Model Performance}
In the ablation experiment section, we further analyze the effects of the Kansformer Encoder, SCConv Encoder, and GLAE components on the classification performance of the SCKansformer model across three datasets. We evaluate the performance of the SCKansformer model by comparing the removal of the Kansformer Encoder, SCConv Encoder, and GLAE components, as well as the performance of the complete SCKansformer model. As illustrated in Table \ref{tab4}, the results indicate that all three modules positively influence the model's classification performance. Specifically:

\textbf{Removal of the Kansformer Encoder:} The Kansformer Encoder replaces the traditional MLP layer via the KAN layer, enhancing the model's ability in nonlinear feature expression and interpretability. Reverting to using the MLP layer in the Transformer Encoder results in a noticeable decline in classification performance. Specifically, Precision drops from 85.82 to 82.27, Recall decreases from 84.28 to 81.23, and F1-score and Accuracy fall from 84.34 and 83.23 to 81.34 and 82.27, respectively. This decline is due to the model's significantly reduced feature expression capability when reverting to the MLP layer, making it insufficient for adequately extracting complex features from cells in microscopic medical images. Consequently, the model learns cell features that are too similar and easily confused among different cell types.
  
\textbf{Removal of the SCConv Encoder:} The SCConv Encoder reduces feature redundancy through the SRU and CRU, thereby enhancing the model's feature expression ability. When the SCConv Encoder is removed, the model's inability to effectively minimize spatial and channel redundancy in microscopic images leads to inadequate feature expression, resulting in significant declines in various metrics. Specifically, Precision drops by 2.03 percentage points, Recall by 1.57 percentage points, F1-score by 3.20 percentage points, and Accuracy by 1.46 percentage points. This underscores the critical importance of the SCConv Encoder in optimizing feature representation in microscopic images, particularly when dealing with their high-dimensional features.
  
\textbf{Removal of the GLAE:} The GLAE combines a MSA module and a Local Part module to capture long-range dependencies and local features of cells in microscopic medical images. When the Global-Local Attention Encoder is removed, the model's ability to capture both global and local features significantly diminishes. The model's performance metrics notably decline, with Precision dropping by 10.44 percentage points, Recall by 8.97 percentage points, F1-score by 9.55 percentage points, and Accuracy by 8.36 percentage points, leading to a marked deterioration in classification performance. This highlights the crucial role of global and local attention mechanisms in the accurate extraction of complex relationships between cells in microscopic image analysis.

\begin{table*}[h]
\centering
\caption{SCKansformer ablation experiments on the private BMCD-FGCD dataset.}
\label{tab4}
\begin{tabularx}{\textwidth}{
    >{\centering\arraybackslash\hsize=0.7\hsize}X
    >{\centering\arraybackslash\hsize=0.5\hsize}X
    >{\centering\arraybackslash\hsize=0.5\hsize}X
    >{\centering\arraybackslash\hsize=0.5\hsize}X
    >{\centering\arraybackslash\hsize=0.5\hsize}X
    >{\centering\arraybackslash\hsize=0.5\hsize}X
  }
\toprule
\textbf{Method} & \textbf{Precision} & \textbf{Recall} & \textbf{F1-score} & \textbf{Accuracy} & \textbf{Balanced Accuracy}\\
\hline
w/o SCConv Encoder & 83.79 & 82.71 & 81.14 & 81.77 & 77.50 \\
w/o Global-Local Attention Encoder & 75.38 & 75.31 & 74.79 & 74.87 & 70.33 \\
w/o Kansformer Encoder KAN layer & 82.27 & 81.23 & 81.34 & 82.37 & 80.47 \\
\textbf{SCKansformer (Ours)} & \textbf{85.82} & \textbf{84.28} & \textbf{84.34} & \textbf{83.23} & \textbf{82.91} \\
\bottomrule
\end{tabularx}
\end{table*}

\subsubsection{Validation of Model Generalizability}
Additionally, we conducted the same ablation experiments on the publicly available leukocyte datasets, PBC and ALL-IDB, which feature balanced categories. These experiments further confirmed the significant impact of each key component on classification accuracy. As shown in Table \ref{tab5} and Table \ref{tab6}, the removal of any component led to a marked decline in classification accuracy, emphasizing the importance of these components in enhancing the model's feature representation capabilities, optimizing feature expression, and effectively capturing both global and local features of microscopic images. Combined with the ablation experiment results on the imbalanced BMCD-FGCD dataset, together they provide comprehensive support for the robustness and generalizability of the SCKansformer model.

\begin{table*}[h]
\centering
\caption{SCKansformer ablation experiments on the PBC public dataset.}
\label{tab5}
\begin{tabularx}{\textwidth}{
    >{\centering\arraybackslash\hsize=0.7\hsize}X
    >{\centering\arraybackslash\hsize=0.5\hsize}X
    >{\centering\arraybackslash\hsize=0.5\hsize}X
    >{\centering\arraybackslash\hsize=0.5\hsize}X
    >{\centering\arraybackslash\hsize=0.5\hsize}X
    >{\centering\arraybackslash\hsize=0.5\hsize}X
  }
\toprule
\textbf{Method} & \textbf{Precision} & \textbf{Recall} & \textbf{F1-score} & \textbf{Accuracy} & \textbf{Balanced Accuracy}\\
\hline
w/o SCConv Encoder & 97.34 & 97.31 & 97.31 & 97.32 & 97.36 \\
w/o Global-Local Attention Encoder & 88.01 & 87.99 & 87.87 & 87.99 & 84.97 \\
w/o Kansformer Encoder KAN layer & 95.13 & 95.14 & 95.14 & 95.13 & 92.70 \\
\textbf{SCKansformer (Ours)} & \textbf{99.47} & \textbf{99.46} & \textbf{99.46} & \textbf{99.46} & \textbf{99.37} \\
\bottomrule
\end{tabularx}
\end{table*}

\begin{table*}[h]
\centering
\caption{SCKansformer ablation experiments on the ALL-IDB public dataset.}
\label{tab6}
\begin{tabularx}{\textwidth}{
    >{\centering\arraybackslash\hsize=0.7\hsize}X
    >{\centering\arraybackslash\hsize=0.5\hsize}X
    >{\centering\arraybackslash\hsize=0.5\hsize}X
    >{\centering\arraybackslash\hsize=0.5\hsize}X
    >{\centering\arraybackslash\hsize=0.5\hsize}X
    >{\centering\arraybackslash\hsize=0.5\hsize}X
  }
\toprule
\textbf{Method} & \textbf{Precision} & \textbf{Recall} & \textbf{F1-score} & \textbf{Accuracy} & \textbf{MCC}\\
\hline
w/o SCConv Encoder & 96.43 & 96.15 & 96.12 & 96.15 & 92.58 \\
w/o Global-Local Attention Encoder & 89.43 & 86.51 & 86.29 & 86.53 & 75.88 \\
w/o Kansformer Encoder KAN layer & 98.15 & 98.08 & 98.08 & 98.08 & 96.23 \\
\textbf{SCKansformer (Ours)} & \textbf{99.90} & \textbf{99.85} & \textbf{99.85} & \textbf{99.86} & \textbf{99.83} \\
\bottomrule
\end{tabularx}
\end{table*}

\section{Conclusion}

In this paper, we propose SCKansformer, a novel fine-grained classification model for bone marrow blood cells that addresses key challenges such as insufficient feature expressiveness, poor interpretability, and feature redundancy in current methods. We evaluate the SCKansformer model against other advanced cell classification models using three datasets: the private BMCD-FGCD dataset and the publicly available PBC and ALL-IDB datasets. The experimental results show that SCKansformer significantly outperforms existing advanced microcell classification methods across these datasets, excelling in all metrics. Additionally, we conduct ablation experiments on the BMCD-FGCD, PBC, and ALL-IDB datasets to determine the importance of key components of the model. The results validate the positive impact of these components on the overall model performance.

In addition, the development of myeloid blood cell classification technology has been significantly constrained by the size and quality of existing datasets. To address this issue, we have collaborated with the Department of Hematology at Zhejiang Hospital to develop the BMCD-FGCD dataset. This dataset contains over 10,000 samples with nearly 40 fine-grained classifications. We have decided to make this private dataset publicly available to the academic community. The establishment and release of the BMCD-FGCD dataset not only provide a robust database for our research but also serve as a valuable resource for advancing the field. We anticipate that the availability of this high-quality dataset will promote further development in myeloid blood cell classification. Future work will focus on expanding dataset diversity and refining the model to ensure its applicability in clinical settings.

\section{References}
\bibliographystyle{ieeetr}
\bibliographystyle{ieeetr}

\end{document}